\documentclass[aps,prb,onecolumn,showpacs,superscriptaddress,floatfix,preprint]{revtex4}

\usepackage[latin1]{inputenc}
\usepackage[T1]{fontenc}
\usepackage{amssymb,amsmath}
\usepackage{times}
\usepackage{graphicx}
\usepackage{wasysym}
\usepackage{subfigure}
\usepackage[sort&compress]{natbib}
\usepackage{dcolumn}
\usepackage{bm}
\usepackage{overpic}
\usepackage{color}
\usepackage{anysize}
\usepackage{multirow}

\begin{document}

\title{Role of Magnetic Exchange Energy on Charge Ordering in \emph{R}$_{1/3}$Sr$_{2/3}$FeO$_{3}$ (\emph{R} = La, Pr, and Nd)}

\author{J. Ma$^*$}
\affiliation{Ames Laboratory, US-DOE, Ames, Iowa 50011, USA}
\affiliation{Department of Physics and Astronomy, Iowa State University, Ames, Iowa 50011, USA}
\author{J.-Q. Yan$^*$}
\affiliation{Ames Laboratory, US-DOE, Ames, Iowa 50011, USA}
\author{S. O. Diallo$^*$}
\affiliation{Ames Laboratory, US-DOE, Ames, Iowa 50011, USA}
\author{R. Stevens}
\affiliation{Los Alamos National Laboratory, Los Alamos, New Mexico 87545, USA}
\author{A. Llobet}
\affiliation{Los Alamos National Laboratory, Los Alamos, New Mexico 87545, USA}
\author{F. Trouw}
\affiliation{Los Alamos National Laboratory, Los Alamos, New Mexico 87545, USA}
\author{D. L. Abernathy}
\affiliation{Oak Ridge National Laboratory, PO Box 2008, Oak Ridge, TN 37831, USA}
\author{M. B. Stone}
\affiliation{Oak Ridge National Laboratory, PO Box 2008, Oak Ridge, TN 37831, USA}
\author{R. J. McQueeney}
\affiliation{Ames Laboratory, US-DOE, Ames, Iowa 50011, USA}
\affiliation{Department of Physics and Astronomy, Iowa State University, Ames, Iowa 50011, USA}
\date{\today}

\begin{abstract}

Inelastic neutron scattering is applied to study the role of magnetism in stabilizing the charge ordered state in \emph{R}$_{1/3}$Sr$_{2/3}$FeO$_{3}$ (\emph{R}SFO) (\emph{R} = La, Pr, and Nd). The ratio of the ferromagnetic exchange energy (\emph{J}$_F$) and antiferromagnetic exchange energy (\emph{J}$_{AF}$), $\mid$\emph{J}$_{F}$/\emph{J}$_{AF}$$\mid$, is a key indicator of the stability of the charge ordered and antiferromagnetic ordered state. This ratio is obtained from the spin wave spectrum by inelastic neutron scattering and is sufficiently large to suggest that the magnetic exchange energy alone can stabilize the charge ordered state in La$_{1/3}$Sr$_{2/3}$FeO$_{3}$ and Pr$_{1/3}$Sr$_{2/3}$FeO$_{3}$. The exchange ratio decreases from La$_{1/3}$Sr$_{2/3}$FeO$_{3}$ to Nd$_{1/3}$Sr$_{2/3}$FeO$_{3}$ which indicates a gradual destabilization of the magnetic exchange mechanism for charge ordering in correspondence with the observed reduction in the ordering temperature.

\end{abstract}


\maketitle

\section*{I) Introduction}

The charge-ordering (CO) transition is often encountered in complex transition-metal oxides (TMO) and has been the focus of intense inquiry and debate in condensed matter science in the past years. The metal-insulator transition that occurs as the temperature decreases across the CO transition temperature, \emph{T}$_{CO}$, is associated with a change from an itinerant electronic state to a more localized state. The CO state plays an important role in various systems, including the superconducting cuprates and the proximity of the superconducting state to a spin/charge stripe ordered state,[1] colossal-magnetoresistive manganites, where the CO states compete with ferromagnetic metallic states,[2] and layered nickelates, which form a small polaron lattice.[3] Therefore, understanding the causes and implications of CO phenomena is significantly important. The CO state is also often closely associated with magnetic and orbital ordering, and it is widely recognized that CO can arise from a variety of competing interactions, most importantly the intersite Coulomb interaction, magnetic exchange energy, and electron-phonon interactions, all of which are strongly dependent on the valence states of neighboring metal ions.

The Sr-doped rare earth ferrite \emph{R}$_{1/3}$Sr$_{2/3}$FeO$_{3}$ (\emph{R}SFO) is an interesting example of a CO system where magnetic exchange energy is thought to play a crucial, if not dominant, role in the stability of the CO state. \emph{R}SFO is a perovskite based crystal where the Fe ion adopts a formal fractional valence of 3.67+. Below \emph{T}$_{CO}$, it has been proposed that charge disproportionation occurs according to 3Fe$^{3.67+}$ $\rightleftharpoons$ 2Fe$^{3+}$ + Fe$^{5+}$, with the different iron valences ordering in planes containing a repeating arrangement of 3+, 3+, 5+ ions perpendicular to the body diagonal [111]$_c$.[4, 5] The CO occurs simultaneously with antiferromagnetic (AF) order. Recently, we used inelastic neutron scattering (INS) measurements of the spin wave spectrum to demonstrate the plausibility that the magnetic exchange energy is the dominant interaction giving rise to CO in La$_{1/3}$Sr$_{2/3}$FeO$_{3}$ (LSFO).[6] Our results show that the observed CO ground state can be stabilized by large ferromagnetic (F) exchange (\emph{J}$_F$) occurring between the nearest-neighbor (\emph{NN}) Fe$^{3+}$-Fe$^{5+}$ pairs. However, this conclusion was made based on the assumption that the intersite Coulomb interaction (i.e. the Madelung energy) is suppressed by strong electronic screening enabled by the small charge transfer (CT) gap (several to 10's of meV) observed in the system.[6] Since we cannot directly determine the contribution of the Coulomb interaction, one possible way to verify the dominance of the magnetic exchange mechanism is to measure the magnetic exchange energies in other \emph{R}SFO (for example, \emph{R} = Pr and Nd) compounds. For the smaller Pr$^{3+}$ and Nd$^{3+}$ ions, increased lattice distortions lead to a larger charge-transfer gap and a narrower electronic bandwidth. This should lead to reduced screening and increase the stability of the CO state due to a greater influence of the Coulomb interaction. However, \emph{T}$_{N}$ and \emph{T}$_{CO}$ are known to be suppressed by smaller \emph{R}$^{3+}$.[5] In 1998, T. Mizokawa and A. Fujimori proposed that the $\mid$\emph{J}$_F$/\emph{J}$_{AF}$$\mid$ exchange ratio is a good indicator of the propagation direction of CO ordering in the limit where magnetic energy is dominant. If the ratio is larger than 1, the charges will be ordered along [111]; if the ratio is less than 1, the charges will be ordered along [100].[7] Thus, if neutron scattering measurements can indicate a weakening of the magnetic exchange ratio, this would give additional support to the magnetic mechanism for CO.

To verify this hypothesis, we study the spin wave spectrum in \emph{R}SFO with different \emph{R}$^{3+}$ ions. Based on INS measurement of the powder sample in principal, the magnetic exchange energies, \emph{J}$_F$ and \emph{J}$_{AF}$, can be obtained,[8] and the related exchange ratio could be calculated. The corresponding contribution of the magnetic energy to the CO state could then be considered. However, the magnetic spectra of PSFO and NSFO are not as simple as that of previously studied LSFO system,[6] where only Fe moments contribute to the neutron magnetic cross-section, as the Pr$^{3+}$ and Nd$^{3+}$ ions themselves possess magnetic moments. The neutron intensity from crystal electronic field (CEF) excitations of the magnetic Pr$^{3+}$ and Nd$^{3+}$ ions makes the extraction of the Fe spin wave spectrum more complicated. To better account for the rare earth CEF excitations, we also investigate the parent compounds, \emph{R}FO (\emph{R} = Pr and Nd). All of the \emph{R}FO parent compounds are insulators with G-type antiferromagnetic ordering of Fe atoms occurring at high temperatures (\emph{T}$_{N}$ $\approx$ 700 K). The rare earth ions in \emph{R}FO are expected to have similar CEF spectra to \emph{R}SFO once the simple dilution of the rare earth site by Sr is taken into account. After accounting for the CEF intensities of the \emph{R}$^{3+}$ ions in the total magnetic cross-section of \emph{R}SFO, the INS data may be compared to calculations of the spin wave spectra and their corresponding cross-sections using a Heisenberg model. We show that the ratio $\mid$ \emph{J}$_F$/\emph{J}$_{AF}$ $\mid$ is found to decrease with smaller \emph{R}$^{3+}$ which can account for the reduction in \emph{T}$_{N}$ by the magnetic mechanism.

\section*{II) Experiment}

Polycrystalline \emph{R}FeO$_3$ (\emph{R}FO) and \emph{R}SFO (\emph{R} = La, Pr, and Nd) were prepared by a conventional solid-state reaction method. Stoichiometric amounts of La$_2$O$_3$, or Nd$_2$O$_3$, or Pr$_6$O$_{11}$, SrCO$_{3}$, and Fe$_2$O$_3$ were mixed by grinding with mortar and pestle. The mixtures were transferred to an Al$_2$O$_3$ crucible and calcined several times in air at temperatures of 1100 $^\circ$C and 1200 $^\circ$C respectively for 24 hours. Then, the press-formed pellets were sintered in air at 1250$^\circ$C and 1350$^\circ$C for 30 hours, respectively. As the ionic size decreases from La to Nd, the \emph{R}SFO compounds tend to be more oxygen deficient.[9] PSFO and NSFO were annealed under oxygen pressure (10 bar) at 600$^\circ$C for 72 hours. Room temperature powder X-ray diffraction (XRD) patterns were performed on a Rigaku Miniflex X-ray diffractometer with Cu \emph{K}$_\alpha$ radiation to confirm phase purity. No impurities were observed. The oxidation state of iron was determined by iodometric titration and is listed in Table I.

\begin{figure*} [tbh]
  \includegraphics[width=1.0\textwidth]{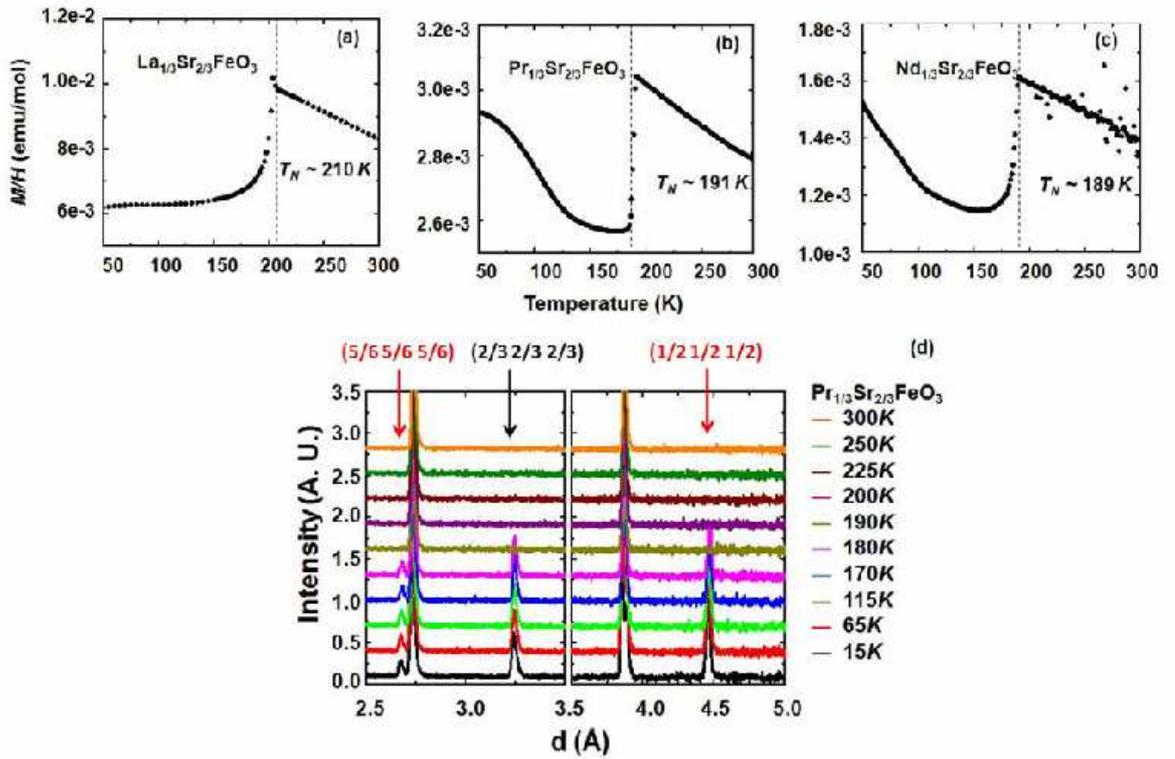}\\
  \caption{(color online) Magnetization of \emph{R}SFO (\emph{R} = La(a), Pr(b) and Nd(c)) as determined by ZFC SQUID measurements; (d) powder neutron diffraction of PSFO from 300 K to 15 K as a function of d-spacing. The arrows show the positions of the magnetic Bragg (red) and charged-ordered superlattice (black) peaks.}
  \label{}
\end{figure*}

The CO and AF transition temperatures for the Sr-doped samples were determined by neutron powder diffraction (NPD), using the High-Intensity Powder Diffractometer (HIPD) at the Lujan Center at Los Alamos National Laboratory, and zero field cooled (ZFC) magnetization measurements using a superconducting quantum interference device (SQUID) magnetometer.  These characterization data are shown in Fig. 1. The \emph{R}SFO systems were found to have identical magnetic propagation vectors and charge-order propagation vectors (inferred via the development of structural superlattice peaks) below \emph{T}$_{N}$. Fig. 1(d) shows temperature dependent neutron diffraction data for PSFO and indicates the charge-order propagation vectors at (n/3, n/3, n/3), and magnetic propagation vectors at (n/6, n/6, n/6).[10] The lattice constants determined from refinement of 300 K NPD and magnetic transition temperatures are listed in Table I. The bond lengths and bond angles determined from NPD patterns at 300 K are listed in Table II. And the geometric tolerance factor (\emph{t}) is expressed as,

\begin{equation}
 t = \frac{<\emph{R}-O>}{\sqrt{2}<Fe-O>} \, ,\label{fiteq}
\end{equation}
where <\emph{R}-\emph{O}> and <\emph{Fe}-\emph{O}> are the average bond lengthes of \emph{R}-\emph{O} and \emph{Fe}-\emph{O}.

INS measurements were performed on the Pharos spectrometer at the Lujan Center of Los Alamos National Laboratory and the ARCS spectrometer at the Spallation Neutron Source (SNS) of Oak Ridge National Laboratory. Both instruments are direct geometry time-of-flight (TOF) spectrometers and measure the scatterring intensity over a wide range of energy transfers ($\hbar$$\omega$) and scattering angles between 1$^\circ$-140$^\circ$, thereby allowing determination of a large swath of the scattering intensity, \emph{S}(Q, $\omega$), as a function of momentum transfer ($\hbar$Q) and energy transfer ($\hbar$$\omega$).

\begin{table*} [tph]
\caption{Lattice and magnetic parameters of \emph{R}FO and \emph{R}SFO (\emph{R} = La, Pr and Nd) as determined by X-ray/neutron scattering measurements at 300 K, SQUID measurements(\emph{T}$_N$), and iodometric titration (oxygen deficiency). }
\begin{tabular}{cccc}
\hline
   & $\,  \,  \,  \,  $ LaFeO$_3$ $\,  \,  \,  \,  $  & $\,  \,  \,  \,  $ PrFeO$_3$  $\,  \,  \,  \,  $  & $\,  \,  \,  \,  $  NdFeO$_3$ $\,  \,  \,  \,  $  \\
\hline
Space group           & \emph{Pnma} & \emph{Pnma} & \emph{Pnma} \\
\hline
Lattice constant(\AA)         & & & \\
$\tiny{a}$   & 5.56 $\pm$ 0.01  & 5.57 $\pm$ 0.01 & 5.59 $\pm$ 0.01 \\
$\tiny{b}$          & 7.85 $\pm$ 0.02 & 7.79 $\pm$ 0.02 & 7.76 $\pm$ 0.02 \\
$\tiny{c}$          & 5.56 $\pm$ 0.01 & 5.48 $\pm$ 0.01 & 5.45 $\pm$ 0.01 \\
\hline
$\,  \,  \,  \,  $ \emph{T}$_N$ (K)[11]          & 738  & 707 & 693 \\
\hline
& $\,  \,  \,  \,  $ La$_{1/3}$Sr$_{2/3}$FeO$_{3}$ $\,  \,  \,  \,  $  & $\,  \,  \,  \,  $ Pr$_{1/3}$Sr$_{2/3}$FeO$_{3}$  $\,  \,  \,  \,  $  & $\,  \,  \,  \,  $  Nd$_{1/3}$Sr$_{2/3}$FeO$_{3}$ $\,  \,  \,  \,  $  \\
\hline
Space group           & \emph{R}$\bar{3}$c & \emph{R}$\bar{3}$c & \emph{R}$\bar{3}$c \\
\hline
Lattice constant(\AA)         & & & \\
$\tiny{a}$ = $\tiny{b}$   & 5.48 $\pm$ 0.01 &  5.48 $\pm$ 0.01 & 5.47 $\pm$ 0.01 \\
$\tiny{c}$          & 13.41 $\pm$ 0.03 & 13.37 $\pm$ 0.03 & 13.34 $\pm$ 0.03 \\
\hline
$\,  \,  \,  \,  $ \emph{T}$_N$ (K)          & 210 $\pm$ 2.0  & 191 $\pm$ 2.0 & 189 $\pm$ 2.0 \\
\hline
$\,  \,  \,  \,  $ Oxygen $\,  $ stoichiometry        & 2.94 $\pm$ 0.03 & 2.97 $\pm$ 0.03 & 2.97 $\pm$ 0.03 \\
\hline

\end{tabular}
\label{Table.I.}
\end{table*}

\begin{table*}[tbh]
\caption{The structural properties of \emph{R}FO and \emph{R}SFO as determined by NPD at 300 K.}

\begin{tabular}{cccc}
\hline
 & LaFeO$_3$  & PrFeO$_3$ & NdFeO$_3$  \\
\hline
Bond length(\AA)$\,  \,   \,  \,  \,  $ & & & \\
 $\,  \,   \,  \,  \,   \,  $ < \emph{R} - O > & 2.694 $\pm$ 0.008 & 2.627 $\pm$ 0.008 & 2.609 $\pm$ 0.008 \\
 $\,  \,   \,  \,  \,   \,  $ Fe - O(1) & 2.002 $\pm$ 0.004 & 2.004 $\pm$ 0.004 & 2.005 $\pm$ 0.004 \\
 $\,  \,   \,  \,  \,   \,  $ Fe - O(2) & 2.004 $\pm$ 0.004 & 2.006 $\pm$ 0.004 & 2.007 $\pm$ 0.004 \\
 $\,  \,   \,  \,  \,   \,  $ Fe - O(2) & 2.005 $\pm$ 0.004 & 2.015 $\pm$ 0.004 & 2.017 $\pm$ 0.004 \\
Bond angle $\,  \,   \,  \,  \,   \,  \,  \,   \,  \,   $ & & & \\
$\,  \,   \,  \,  \,   \,  $$\angle$ Fe - O(1) - Fe & 157.6$^\circ$ $\pm$ 0.3$^\circ$ & 153.3$^\circ$ $\pm$ 0.3$^\circ$ & 151.2$^\circ$ $\pm$ 0.3$^\circ$ \\
$\,  \,   \,  \,  \,   \,  $$\angle$ Fe - O(2) - Fe & 157.5$^\circ$ $\pm$ 0.3$^\circ$ & 152.4$^\circ$ $\pm$ 0.3$^\circ$ & 151.4$^\circ$ $\pm$ 0.3$^\circ$ \\
Geometric tolerance factor $\,  \,   \,  \,  \,   \,  \,  \,   \,  \,   $ & & & \\
 $\,  \,   \,  \,  \,   \,  $\emph{t} & 0.951 $\pm$ 0.001 & 0.925 $\pm$ 0.001 & 0.918 $\pm$ 0.001 \\
\hline
 & La$_{1/3}$Sr$_{2/3}$FeO$_3$  $\,  \,   \,  \,  \,   \,  $& Pr$_{1/3}$Sr$_{2/3}$FeO$_3$ $\,  \,   \,  \,  \,   \,  $& Nd$_{1/3}$Sr$_{2/3}$FeO$_3$  $\,  \,   \,  \,  \,   \,  $\\
\hline
Bond length(\AA)$\,  \,   \,  \,  \,   \,  \,  \,   \,  $ & & & \\
 $\,  \,   \,  \,  \,   \,  $\emph{R} - O & 2.7413 $\pm$ 0.006 & 2.7377 $\pm$ 0.006 & 2.7329 $\pm$ 0.006 \\
 $\,  \,   \,  \,  \,   \,  $Fe - O & 1.940 $\pm$ 0.004 & 1.941 $\pm$ 0.004 & 1.939 $\pm$ 0.004 \\
Bond angle $\,  \,   \,  \,  \,   \,  \,  \,   \,  \,   $ & & & \\
$\,  \,   \,  \,  \,   \,  $ $\angle$ Fe - O - Fe & 173.2$^\circ$ $\pm$ 0.4$^\circ$ & 170.5$^\circ$ $\pm$ 0.3$^\circ$ & 169.3$^\circ$ $\pm$ 0.3$^\circ$ \\
Geometric tolerance factor $\,  \,   \,  \,  \,   \,  \,  \,   \,  \,   $ & & & \\
 $\,  \,   \,  \,  \,   \,  $\emph{t} & 0.999 $\pm$ 0.002 & 0.997 $\pm$ 0.002 & 0.996 $\pm$ 0.002 \\
\hline
\end{tabular}
\end{table*}

On ARCS, powders ($\sim$14 g) of \emph{R}FO and \emph{R}SFO (\emph{R} = La, Pr, and Nd) were packed in 5 aluminum foil sachets. The sachets were placed in an aluminum can filled with He exchange gas whose size was approximately 4.5 cm $\times$ 6.5 cm $\times$ 0.5 cm. INS spectra were measured with an incident energy (\emph{E}$_i$) of 180 meV. On Pharos, $\sim$ 50 g of \emph{R}FO and \emph{R}SFO (\emph{R} = La, and Nd) were packed in a flat aluminum can (6 cm $\times$ 6 cm $\times$ 0.5 cm), and \emph{E}$_i$s were 120 meV and 160 meV. The face of the sample can was oriented at 135$^\circ$ to the incident neutron beam for both instruments in a transmission geometry. To achieve adequate statistics, the sample was measured for approximately $\sim$ 24 hours on Pharos, and $\sim$ 5 hours on ARCS. Empty sample can measurements were also performed and subtracted from the data presented.

\section*{III) Results and Discussion}

The unpolarized inelastic neutron scattering cross-section contains contributions from both magnetic and phonon scattering. In order to isolate the spin wave spectrum, the magnetic scattering must first be separated from the phonon scattering. This is accomplished by using the fact that the magnetic scattering intensity decreases with Q (or 2$\theta$) due to the magnetic form factor, while phonon scattering intensity increases proportional to Q$^2$.

\begin{figure*} [tbh]
  \includegraphics[width=1.1\textwidth]{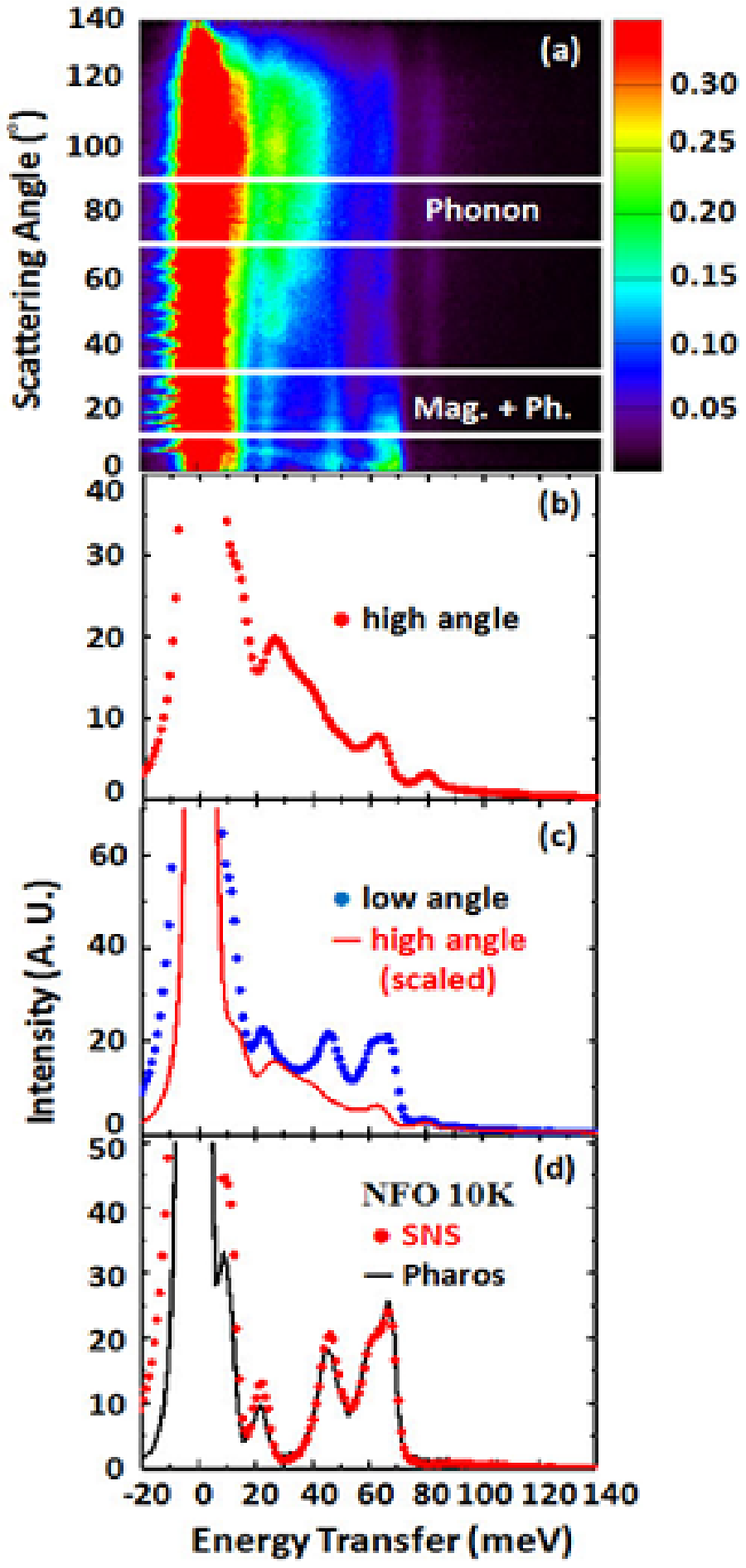}\\
  \caption{(color online) (a) Inelastic neutron scattering intensity (\emph{E}$_i$ = 180 meV) of NFO (color scale) versus scattering angle and energy transfer at \emph{T} = 10 K as measured at ARCS. Horizontal white lines delineate regions where phonon and magnetic scattering are isolated; (b) neutron intensity summed over the angular range from 75-95$^\circ$ originating from phonons (red dots); (c) neutron intensity summed over the low angular range from 10-30$^\circ$ (blue dots) and phonon background scaled from the high angle sum; (d) the isolated magnetic scattering of ARCS data (red dots) and Pharos data(black line).}
  \label{}
\end{figure*}

The INS data of NFO taken on ARCS at \emph{T} = 10 K with \emph{E}$_i$ = 180 meV are shown in Fig. 2(a) and used as an example of the data treatment. A similar analysis was performed for LFO on Pharos as outlined in Ref. [6]. The data summed over the high angle range of 2$\theta$ = 75-95$^\circ$ contain primarily phonon scattering, Fig. 2(b), while the data within the low angle range of 10-30$^\circ$ contain scattering from both phonons and spin waves arising from the G-type AFM order of Fe$^{3+}$ and the CEF of Nd$^{3+}$, Fig. 2(c). The magnetic scattering in NFO was isolated by subtracting the high angle phonon data from low angle data after scaling the high angle data by a constant factor. A comparison of the scaled high angle data to the low angle data is shown in Fig. 2(c) and the phonon subtracted data is shown in Fig. 2(d). In order to compare the two instruments used for the INS measurements, the magnetic spectrum of NFO measured on the Pharos is overplotted in Fig. 2(d). The spectrum agrees with each other very well.

\section*{A) Magnetic spectra of \emph{R}FO}

\section*{a) Fe Spin Waves}

We now discuss the analysis of collective Fe spin waves below \emph{T}$_{N}$ in the parent \emph{R}FO compounds. In \emph{R}FO, \emph{NN} Fe$^{3+}$(3\emph{d}$^5$) spins are coupled by strong \emph{AF} superexchange interactions (\emph{J}$_{AF}$ $<$ 0). According to the single-crystal INS studies of TmFeO$_3$,[12] the spin waves can be approximated using a Heisenberg model Hamiltonian with only isotropic \emph{NN} exchange interaction,

\begin{equation}
 {\bf{H}} = - J_{AF} \sum_{<i, j>} {\bf{S}}_i \cdot {\bf{S}}_j\, ,\label{fiteq}
\end{equation}
where ${\bf{S}}_i$ and ${\bf{S}}_j$ represent the spin vectors of the \emph{i}th and \emph{j}th iron atoms that are \emph{NN}s.

The INS information obtained from the polycrystalline samples is related to the spin wave density-of-states (SWDOS) via a powder-averaging over all crystallographic directions.[8] For the G-type LFO spin waves, the SWDOS consists of a single sharp peak at an energy of 6$\mid$\emph{J}$_{AF} \mid$ \emph{S}$^{3+}$ and \emph{S}$^{3+}$ is the spin magnitude of Fe$^{3+}$ ion. Assuming \emph{S}$^{3+}$ = 5/2, a value of \emph{J}$_{AF}$ = -4.9 meV can be determined from the position of this single peak at 73 meV, shown in Fig. 3. The value of \emph{J}$_{AF}$ is also listed in Table III.

\begin{figure*} [tbh]
  \includegraphics[width=0.6\textwidth]{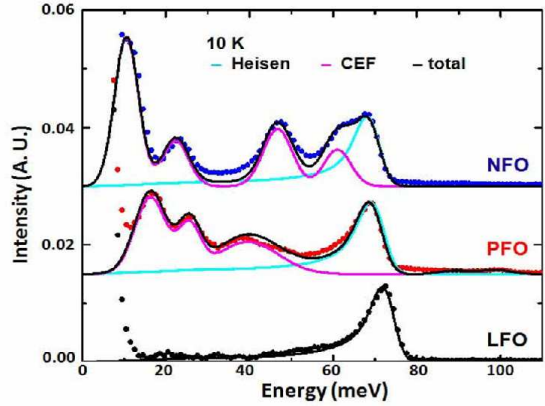}\\
  \caption{(color online) The experimental angle-averaged magnetic INS data for \emph{R}FO (\emph{R} = La, Pr and Nd) at 10K (dots). Heisenberg model calculation of the powder-averaged Heisenberg spin waves cross-section (\emph{Heisen}) for G-type magnetic order are shown as solid cyan lines. Additional peaks in the spectra for \emph{R} = Pr and Nd and associated solid pink lines are the fits of CEF excitations.}
  \label{}
\end{figure*}

Figure 3 shows the phonon subtracted magnetic data for LFO, PFO, and NFO found via the procedure described in Fig. 2. As La$^{3+}$ has no f-electrons, there are no CEF excitations existing in LFO. While PFO and NFO have similar G-type magnetic structure as LFO, it is clear that the neutron spectra of PFO and NFO contain addiontal magnetic excitations due to \emph{R}$^{3+}$ CEF excitations which will be discussed later. In order to determine which of the observed peaks are due to Fe spin waves and which are CEFs, we can use mean-field theory and knowledge of \emph{T}$_{N}$ to estimate \emph{J}$_{AF}$ exchange coupling between Fe ions for \emph{R} = Nd and Pr. In mean-field theory,[13]
\begin{equation}
 3 k_B T_N = N \mid J_{AF} \mid \sqrt{S(S+1)}\, ,\label{fiteq}
\end{equation}
where \emph{N} = 6 and \emph{S} is the spin angular momentum, 5/2.[8] With this assumption, the magnetic exchange energies are expected to weaken slightly upon going from LFO to NFO due to the decrease of \emph{T}$_{N}$. However, it is well-known that \emph{T}$_{N}$ is overestimated in mean field theory and therefore the mean-field estimate is smaller than that observed by neutron scattering. Table III lists the AF exchange energy based on Eq. (3). A more accurate value of the exchange can then be obtained by fitting the SWDOS peak in the INS data by Heisenberg model, Eq. (2). The results are listed Table III and fits to the SWDOS are shown in Fig. 3. According to superexchange theory, crystalline distortions caused by the smaller rare earth ions result in a bending of the Fe-O-Fe bond angle that weakens \emph{J}$_{AF}$, Table II.[14]

\begin{table*} [tph]
\caption{Magnetic exchange energy of \emph{R}FO (\emph{R} = La, Pr and Nd). }
\begin{tabular}{cccc}
\hline
\emph{J}$_{AF}$ (meV)   & $\,  \,  \,  \,  $ LaFeO$_3$ $\,  \,  \,  \,  $  & $\,  \,  \,  \,  $ PrFeO$_3$  $\,  \,  \,  \,  $  & $\,  \,  \,  \,  $  NdFeO$_3$ $\,  \,  \,  \,  $  \\
\hline
Mean Field           & -3.74 & -3.58 & -3.51 \\
\hline
Heisenberg model           & \multirow{2}{*}{-4.9} & \multirow{2}{*}{-4.69} & \multirow{2}{*}{-4.60} \\
(based on LFO)  \\
\hline
Heisenberg model (fit)           & -4.9 & -4.55 & -4.45 \\
\hline
\end{tabular}
\label{Table.II.}
\end{table*}

\section*{b) Crystal Electric Fields of Pr$^{3+}$ and Nd$^{3+}$}

Mean field theory helps us locate the characteristic energy of spin-wave excitations from the Fe$^{3+}$ ions. The remaining excitations in the phonon subtracted data should be from CEF of \emph{R}$^{3+}$ ions. The cross section for CEF excitations can be written as,[15]

\begin{equation}
\frac{d^2 \sigma}{d \Omega d \omega } \propto [g_J F({\bf{Q}})]^2 e^{-2W} \frac{k^\prime}{k} S_{CEF}({\bf{Q}}, \omega)\, ,
\label{fiteq}
\end{equation}
where ${\bf{Q}}$ = $\vec{k^\prime}$ - $\vec{k}$ is the scattering vector, and $\hbar \omega$ is the energy transfer. \emph{g}$_J$ is Lande factor of the \emph{R}$^{3+}$ ion, \emph{F}(${\bf{Q}}$) is the \emph{R}$^{3+}$ magnetic form factor, \emph{k}$^\prime$ and \emph{k} are values of the initial and final neutron wavevectors, and e$^{-2W}$ is the \emph{R}$^{3+}$ Debye-Waller factor. \emph{S}$_{CEF}$(${\bf{Q}}$, $\omega$) is the response function of the system which is determined by the temperature, CEF level splitting, CEF eigenstates and corresponding CEF matrix elements.

\begin{equation}
S_{CEF}({\bf{Q}}, \omega) = \sum_{i, j} \rho_i \mid \langle i \mid {\bf{J}}_\bot \mid j \rangle \mid ^2 \delta (E_i - E_j - \hbar \omega ) \, ,
\label{fiteq}
\end{equation}
where $\mid$ \emph{i} $\rangle$ and $\mid$ \emph{j} $\rangle$ are the initial and final CEF eigenstates of the system with level energies \emph{E}$_i$ and \emph{E}$_j$. ${\bf{J}}_\bot$ is the component of the total angular momentum operator perpendicular to the scattering vector; $\rho_i$ is the thermal population factor of the initial state. Observable excitations occur between levels which have non-zero matrix elements.

For rare earth ions, the spin-orbit coupling is usually stronger than the CEF potential, and the total angular momentum ${\bf{J}}$ = ${\bf{L}}$ + ${\bf{S}}$, remains a good quantum number. Therefore, the magnetic form factor in Eq. (4) is given by,[15, 16]

\begin{equation}
F({\bf{Q}}) = \langle j_0(Q) \rangle + \langle j_2(Q) \rangle \frac{J(J+1) + L(L+1) - S(S-1)}{3J(J+1) +  S(S-1) - L(L+1)} \, ,
\label{fiteq}
\end{equation}
where $\langle$ \emph{j}$_0 \rangle$ and $\langle$ \emph{j}$_2 \rangle$ are \emph{Q}-dependent functions whose values are tabulated.[17]

\section*{b.1) NdFeO$_3$}

The ground state of the free Nd$^{3+}$ ion has 3 unpaired \emph{f}-electrons, and the Russel-Saunders term symbol is $^4$\emph{I}$_{9/2}$ with a degeneracy of 2 \emph{J} + 1 = 10. At a point of orthorhombic symmetry in the distorted perovskite cell, the ground state multiplet splits into (2 \emph{J} + 1)/2 = 5 CEF doublets.[16] At 10 K, we observe four CEF transitions at $\sim$ 9, 21, 46 and 60 meV as shown in Fig. 3. These are consistent with previous work,[18, 19] and are associated with excitations from the CEF ground state to each of the four excited states. Due to the possible overlap of CEF excitations with spin waves, the phonon scattering contribution and multiple scattering, the separation of the CEF contribution was done by examining both the $\omega$- and Q- dependence of the total cross-section. The positions and integrated intensities of the Nd$^{3+}$ CEFs in NFO at 10 K were determined by,

\begin{equation}
\emph{S}(Q, \omega) = \emph{S}_{mag}(Q, \omega) + \emph{S}_{phonon}(Q, \omega) + \emph{S}_{CEF} (Q, \omega) + \emph{S}_{multi} (Q, \omega) \, ,
\label{fiteq}
\end{equation}
where \emph{S}$_{mag}$(Q, $\omega$) is the polycrystalline averaged spin wave scattering of Fe$^{3+}$ ions, \emph{S}$_{phonon}$(Q, $\omega$) is the polycrystalline-averaged phonon background, and \emph{S}$_{CEF}$ (Q, $\omega$) is CEF scattering from Nd$^{3+}$ ions.

For simplicity, we treat the phonon scattering from a powder sample in the incoherent approximation. In this approximation, the one-phonon scattering is proportional to the phonon density-of-states (DOS) and can be expressed as,

\begin{equation}
S_{inc, \pm 1-phonon} = \frac{Q^2}{\hbar \omega} Z(\hbar \omega) \langle n+1 \rangle \, , \label{fiteq}
\end{equation}
where Z($\hbar$ $\omega$) is the sum of weighted partial phonon DOS, Z$_i$($\hbar$ $\omega$),
\begin{equation}
Z(\hbar \omega) = \sum_i \frac{b_i^2}{2M_i} e^{-2W_i}Z_i(\hbar \omega) \, , \label{fiteq}
\end{equation}
and $\langle$n+1$\rangle$ is the Bose population factor,
\begin{equation}
\langle n+1 \rangle = \frac{1}{2}[coth(\frac{1}{2} \hbar \omega \beta) + 1] \, , \label{fiteq}
\end{equation}

\begin{figure*} [tbh]
  \includegraphics[width=0.7\textwidth]{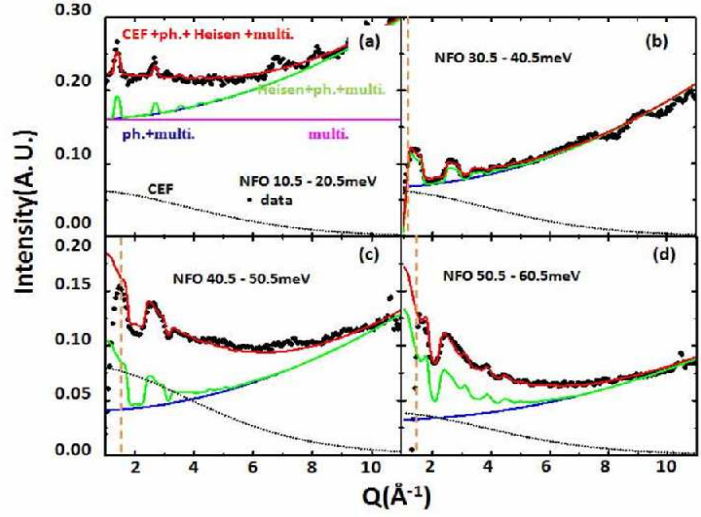}\\
  \caption{(color online) The Q-dependence of the neutron scattering data averaged over different energy transfer ranges in NFO at \emph{T} = 10 K. (a) 10.5-20.5 meV, (b) 30.5-40.5 meV, (c) 40.5-50.5 meV, and (d) 50.5-60.5 meV. The black dots are the experimental data. The blue line is an estimate of the incoherent phonon background (\emph{ph.}) plus multiple scattering (\emph{multi.}). The green line is the calculation of the polycrystalline averaged spin wave scattering (\emph{Heisen}) plus background using the parameters in the text. The red line is sum of the calculation of CEF, the polycrystalline averaged spin wave scattering, background, and multiple scattering. The vertical brown dash line in (b) - (d) is the kinematic limit for neutron data. The dotted line is the fitted CEF excitations.}
  \label{}
\end{figure*}

Constant energy scan obtained from averaging over different energy transfer ranges are shown in Fig. 4. Figure 4(a) is averaged over an energy range from 10.5-20.5 meV and described here as an example. The multiphonon contribution and other background contributions (multi.) are treated as a constant background (horizontal line). The single phonon (ph.) cross-section is determined by assuming an incoherent quadratic Q-dependence (parabola curve) given by Eq. (8). We note that the peaks seen at high-Q (above 5 \AA$^{-1}$) arise from coherent phonon scattering that is not included in the analysis. The spin wave scattering intensities (Heisen) of Fe$^{3+}$ are calculated from a Heisenberg model (zigzag curve). Note that the sharp peaks in the spin wave contribution arise from the coherent scattering, which is included in the Heisenberg model calculations of the powder averaged cross-section. Especially at low energies, the sharp inelastic peaks are coincident with the position of magnetic Bragg peaks. The remaining signal is associated with CEF scattering, which follows the magnetic form factor. The energy ranges where there is a large difference between total fitting line (multiphonon  + phonon + spin wave) and the data signals the presence of a CEF excitation. The dotted line in Fig. 4 is the fitted CEF excitations.

\begin{table*} [tbh]
\caption{The measured and calculated CEF transition energies of Nd$^{3+}$ in NFO at 10 K.[18, 19]}

\begin{tabular}{ccccc}
\hline
\multirow{2}{*}{$\,  \,  \,  \,  \,   \, $Energy levels $\,  \,  \,  \,  \,   \, $}  & \multicolumn{2}{c}{$\hbar \omega$ (meV)} & \multicolumn{2}{c}{Integrated Intensities} \\ \cline{2-5}
 & calculated & measured& calculated & measured \\
\hline
0           & 0.0 & 0.0 & & \\
1           & 10.1 & 9.4 $\pm$ 0.1 & 100.0 & 100.0 $\pm$ 3.1\\
2           & 22.4 & 21.2 $\pm$ 0.1 & 29.3 & 30.9 $\pm$ 1.1\\
3           & 44.7 & 45.7 $\pm$ 0.1 & 46.1 & 41.8 $\pm$ 1.6\\
4           & 60.8 & 59.9 $\pm$ 0.1 & 7.0 & 17.8 $\pm$ 1.3\\
\hline
\end{tabular}
\label{Table.VI.}
\end{table*}

The measured and calculated energies and transition intensities of the CEF levels are listed in Table IV and plotted in Figs. 3 and 4. The CEF excitations of NFO can be found via the single-particle crystal field theory, and is discussed in Refs. [18, 19]. The calculation and measurement of the CEF transitions agree with each other very well. If we normalize the intensities of the CEF transitions to the intensity of the excitation from the ground state to the first excited state, which are 9.4 and 10.1 meV for the measurement and calculation respectively, the measured intensities of the second ( $\sim$ 21.2 meV) and third excited states ($\sim$ 45.7 meV) agree with the calculations based on estimates of the corresponding matrix elements. A comparison of the intensity of the 59.9 meV is more difficult due to the proximity to the magnetic signal from Fe$^{3+}$ spin waves.

\section*{b.2) PrFeO$_3$}

\begin{figure*} [tbh]
  \includegraphics[width=0.90\textwidth]{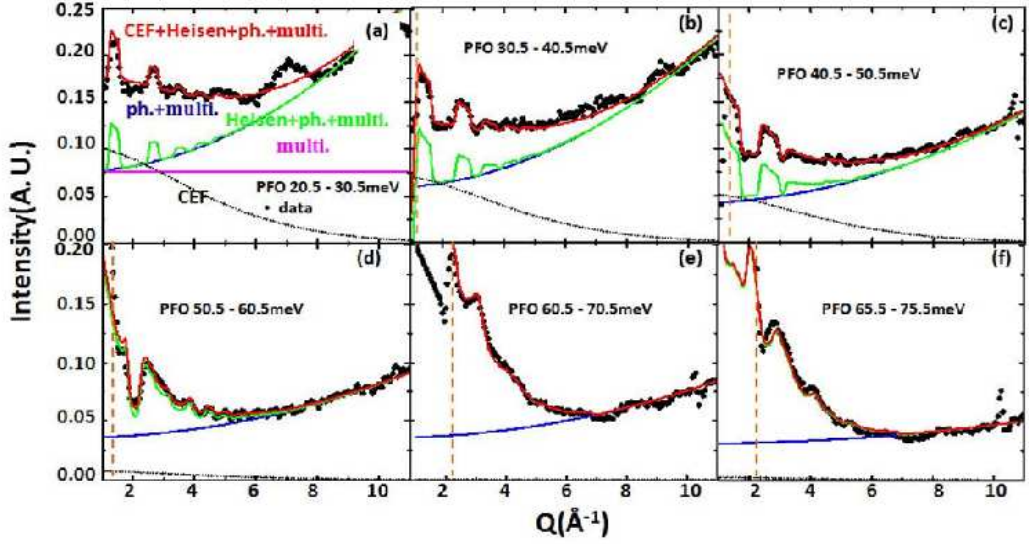}\\
  \caption{(color online) The Q-dependence of the neutron scattering data for different energy transfer ranges (Q-cuts) in PFO at \emph{T} = 10 K. The brown dash line in some different energy transfer ranges ((c), (e), and (f)) indicates the usable low-angle data limit of the experiment. The dotted line is the fitted CEF excitations.}
  \label{}
\end{figure*}

We next examine the CEF transitions in PFO. Although many experimental and theoretical studies on Pr$^{3+}$ in perovskite oxides have been performed,[20, 21] we know of no complete set of experimental data for CEF levels of Pr$^{3+}$ ions in the perovskite structure ABO$_3$. There are 2 unpaired \emph{f}-electrons for Pr$^{3+}$ and the electronic term is $^3$\emph{H}$_4$. Since the structure of PFO is orthorhombically distorted and \emph{J} = 4, the ground state multiplet splits into 2\emph{J} + 1 = 9 singlets. The dipole allowed transitions from the ground state are listed in Table V. Similar to the analysis of NFO, the magnetic form factor of Pr$^{3+}$ in PFO is taken from the literature[17] and used with the previously observed transitions from Ref. [20] in Table V to establish the fraction of different contributions to the total cross-section. Fig. 5 shows the result of the process. The data agree well with the model calculation over a range of wave vector and energy transfers.

\begin{table*} [tbh]
\caption{The measured and calculated CEF transition energies of Pr$^{3+}$ in PFO at 10 K.}

\begin{tabular}{ccccc}
\hline
\multirow{2}{*}{$\,  \,  \,  $Energy levels $\,  \,  \,  $}& \multicolumn{2}{c}{$\,  \,  \,  $PrFeO$_3$ $\,  \,  \,  \,  \,  $} & PrGaO$_3$ & $\,  \,  \,  \,  \,  $ Integrated Intensities $\,  \,  \,  $ \\
& \multicolumn{2}{c}{$\,  \,  \,  $(measured) $\,  \,  \,  $} & (calculated)& (present work) \\
\hline
& Ref. [20] & present work & Ref. [21]& \\ \cline{2-5}
0           & 0.0 & 0.0 & 0.0 & 0.0 \\
1           & $\,  \,  \,  \,  \,   \, $ 2.0 $\pm$ 0.1 $\,  \,  \,  \,  \,   \, $  & -- -- & 5.6 & -- -- \\
2           & 14.7 $\pm$ 0.4 & 15.2 $\pm$ 0.1  & 16.0 & 100.0 $\pm$ 3.1 \\
3           & 23.2 $\pm$ 0.5 & 24.7 $\pm$ 0.1  & 23.4 & 41.1 $\pm$ 1.6 \\
4           & 36.0 $\pm$ 1.0 & 36.6 $\pm$ 0.5  & 32.9 & 82.2 $\pm$ 2.3 \\
5           & 58.0 $\pm$ 2.0 & -- -- & 67.4 & -- -- \\
6           & -- -- & -- --  & 69.3 & -- -- \\
7           & -- -- & 80.1 $\pm$ 0.5 & 89.5 & 4.2 $\pm$ 0.8 \\
8           & -- -- & 97.9 $\pm$ 0.3 & 113.1 & 4.9 $\pm$ 0.5 \\
\hline
\end{tabular}
\label{Table. V.}

\end{table*}

The CEF levels in PFO were compared to those of Pr$^{3+}$ in PrGaO$_3$ (PGO) because it had a similar Pr$^{3+}$ CEF scattering as PFO and was the only theoretical calculation on CEF of Pr$^{3+}$ in perovskite oxides with the space group \emph{Pnma}.[20, 21] There are discrepancies between our measurements of CEF excitations in PFO and the calculations for PGO. First, the CEF transition predicted to appear around $\sim$ 67 meV was not observed in our measurement. This is likely due to the proximity of this transition to the Fe$^{3+}$ spin wave band. The proposed observation of this CEF transition was mentioned in regards to INS measurements of PFO and PGO.[20, 21]. However, this analysis did not report the spin-wave excitation of Fe$^{3+}$ ions,[21] and the magnetic peak claimed to be a CEF around $\sim$ 60 meV is much more likely to be Fe$^{3+}$ spin waves scattering. An examination of Fig. 5(e) shows that the 67 meV CEF transition may be either too weak to observe (as compared to the Fe$^{3+}$ spin waves) or shifted in energy as compared to PGO. We also observe a weak excitation peak around $\sim$ 100 meV, Fig. 3. This is presumably a CEF excitation, however the matrix element is predicted to be zero.[20]

\section*{B) Magnetic excitations in \emph{R}SFO}

We now take our knowledge of the different scattering cross-sections in the parent \emph{R}FO compounds and use it in an attempt to isolate the scattering arising from the Fe$^{3+}$ spin waves in the doped \emph{R}SFO compounds. Figure 6(a) shows the extracted low-angle magnetic intensity of \emph{R}SFO (\emph{R} = La, Nd, and Pr). In general, we find magnetic signals up to a maximum energy of 120 meV. The high energy portion of the magnetic excitation spectrum between 90 $\sim$ 120 meV is very similar for each compound. In LSFO, this high energy band is associated with spin waves propagating in ferromagnetic channels in the magnetic lattice. The low energy features are more difficult to compare due to the presence of \emph{R}$^{3+}$ CEF excitations.

\begin{figure*} [tbh]
  \includegraphics[width=1.1\textwidth]{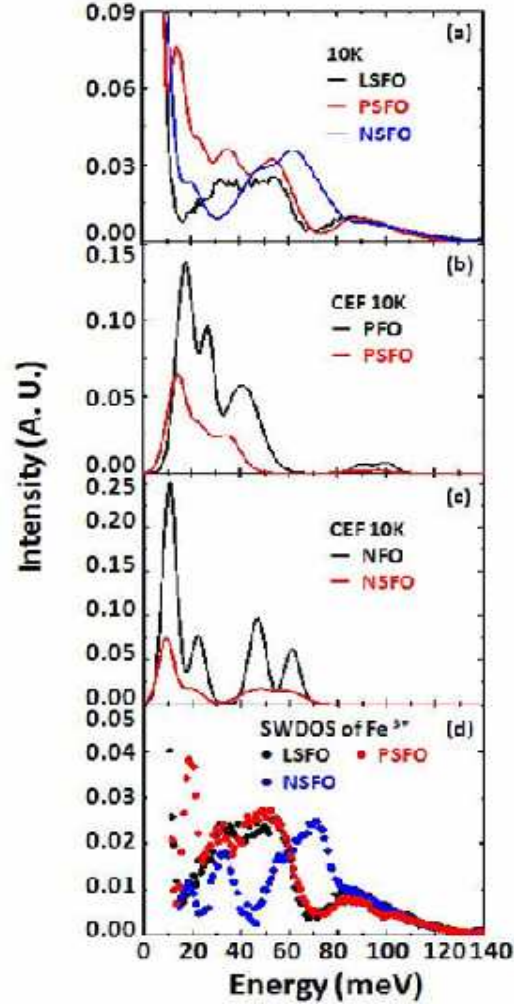}\\
  \caption{(color online) (a) Magnetic inelastic neutron scattering intensity of \emph{R}SFO (\emph{R} = La, Pr, and Nd) versus energy transfer at \emph{T} = 10 K and \emph{E}$_i$ = 180 meV. The calculated CEF intensities of (b) PFO (black line) and PSFO (red line); (c) NFO (black line) and NSFO (red line). (d) The spin wave scattering from Fe$^{3+}$ ions in \emph{R}SFO (\emph{R} = La (black), Pr (red), and Nd (blue))found via a difference of the total magnetic signal (a) minus the fitted CEF contribution of (b) and (c) respectively.}
  \label{}
\end{figure*}

\section*{a) Crystal Electric Fields of Pr$^{3+}$ and Nd$^{3+}$}

The CEF information obtained from the parent \emph{R}FO compounds (\emph{R} = Pr and Nd) can be used as a guide to estimate the CEF contribution in Sr-doped \emph{R}SFO. Because the ionic radius of Sr$^{2+}$ is close to the radii of the \emph{R}$^{3+}$ ion, we assume that the average structural environment of the \emph{R}$^{3+}$-site does not change significantly in \emph{R}SFO with Sr$^{2+}$ doping. Thus, the CEF of \emph{R}$^{3+}$ in \emph{R}SFO would be similar to that of \emph{R}FO. Since 2/3 of \emph{R}$^{3+}$ has been substituted by nonmagnetic Sr$^{2+}$ ions, the integrated intensity of the CEF excitations in \emph{R}SFO should be 1/3 of that found for \emph{R}FO. However, the effect of disorder and other lattice distortions arising from the Sr$^{2+}$ substitution may shift the position and broaden the width of CEF excitations.

\begin{table*}[tbh]
\caption{The transition energies and integrated intensities (area) of Pr$^{3+}$ CEF transitions in PFO and PSFO at 10 K. The last column is a ratio of the integrated intensities of the related excitations.}

\begin{tabular}{cccccc}
\hline
\multirow{2}{*}{E$_{j0}$}   & \multicolumn{2}{c}{PFO}  & \multicolumn{2}{c}{PSFO} & $\,  \,  \,$ \multirow{2}{*}{PFO/PSFO} $\,  \,  \,$ \\ \cline{2-5}
& $\hbar \omega$ (meV) & $\,  \,  \,  \,  \,  \,  \,  $area $\,  \,  \,  \,  \,  \,  \,  $& $\hbar \omega$ (meV) & $\,  \,  \,  \,  \,  \,  \,  $area $\,  \,  \,  \,  \,  \,  \,  $ & \\
\hline
$\,  \, \,  \,   \,  $ E$_{10}$ $\,  \, \,  \,   \,  $       & -- --   & -- -- & -- -- & -- -- & -- -- \\
\hline
E$_{20}$            & 15.2 $\pm$ 0.1 & 1.29 $\pm$ 0.04 & 13.9 $\pm$ 0.3 & 0.68 $\pm$ 0.01 & 1.9 $\pm$ 0.1 \\
\hline
E$_{30}$            & 24.6 $\pm$ 0.1 & 0.53 $\pm$ 0.02 & 23.3 $\pm$ 0.2 & 0.19 $\pm$ 0.01 & 2.9 $\pm$ 0.2 \\
\hline
E$_{40}$            & 38.6 $\pm$ 0.5 & 1.06 $\pm$ 0.03 & 34.1 $\pm$ 0.1 & 0.36 $\pm$ 0.01 & 2.9 $\pm$ 0.1 \\
\hline
E$_{50}$            & -- -- & -- --  & -- -- & -- -- & -- -- \\
\hline
E$_{60}$            & -- -- & -- --  & -- -- & -- -- & -- -- \\
\hline
E$_{70}$            & 88.1 $\pm$ 0.5 & 0.054 $\pm$ 0.01 & 85.9 $\pm$ 0.4 & 0.018 $\pm$ 0.003 & 3.0 $\pm$ 0.4 \\
\hline
E$_{80}$            & 97.9 $\pm$ 0.3 & 0.063 $\pm$ 0.07 & 96.9 $\pm$ 0.4 & 0.021 $\pm$ 0.002 & 3.0 $\pm$ 0.4 \\
\hline
\end{tabular}
\label{Table. VI.}
\end{table*}

The CEF scattering intensities of \emph{R}SFO were fitted with the same method as used for the \emph{R}FO samples. The intensities were initially constrained to be one-third of the corresponding transition in \emph{R}FO and then allowed to vary in the final fits. The fitted results of \emph{R}$^{3+}$ CEFs in the compounds \emph{R}SFO are compared with the data of \emph{R}FO in Fig. 6(b) and (c). The integrated intensities of \emph{R}$^{3+}$ CFE states in these compounds are listed in Table VI (Pr$^{3+}$) and Table VII (Nd$^{3+}$), respectively. The transition energy between different energy levels is defined as \emph{E}$_{ji}$, which \emph{i}th and \emph{j}th are the numbers of the energy levels.

The integrated intensity ratio of the CEF excitations of Nd$^{3+}$ in NFO is consistently 3 times that of the CEFs measured in NSFO as shown in Table VII. For PSFO, the ratio of Pr$^{3+}$ CEF intensities deviates from 3 for the \emph{E}$_{20}$ transition, which overlaps the strong elastic peak.

\begin{table*}[tbh]
\caption{The transition energies and integrated intensities (area) of Nd$^{3+}$ CEF transitions in NFO and NSFO at 10 K. The last column is a ratio of the integrate intensities of these excitations.}

\begin{tabular}{cccccc}
\hline
\multirow{2}{*}{E$_{j0}$}   & \multicolumn{2}{c}{NFO}  & \multicolumn{2}{c}{NSFO} & $\,  \,  \,$ \multirow{2}{*}{NFO/NSFO} $\,  \,  \,$ \\ \cline{2-5}
& \emph{E} (meV) & $\,  \,  \,  \,  \,  \,  \,  $area $\,  \,  \,  \,  \,  \,  \,  $& \emph{E} (meV) & $\,  \,  \,  \,  \,  \,  \,  $area $\,  \,  \,  \,  \,  \,  \,  $ &\\
\hline
$\,  \, \,  \,   \,  $ E$_{10}$ $\,  \, \,  \,   \,  $           & 9.1 $\pm$ 0.1 & 1.8 $\pm$ 0.1 & 9.1 $\pm$ 0.4 & 0.6 $\pm$ 0.1 & 3.0 $\pm$ 0.3\\
\hline
E$_{20}$            & 20.8 $\pm$ 0.1 & 0.6 $\pm$ 0.1 & 19.2 $\pm$ 0.2 & 0.2 $\pm$ 0.1 & 3.0 $\pm$ 0.4\\
\hline
E$_{30}$            & 45.2 $\pm$ 0.1 & 0.9 $\pm$ 0.1 & 46.9 $\pm$ 0.4 & 0.3 $\pm$ 0.1 & 3.0 $\pm$ 0.5\\
\hline
E$_{40}$            & 59.5 $\pm$ 0.2 & 0.5 $\pm$ 0.1 & 60.9 $\pm$ 0.4 & 0.2 $\pm$ 0.1 & 3.1 $\pm$ 0.8\\
\hline
\end{tabular}
\end{table*}

\section*{b) Fe Spin Wave Excitations}

After subtracting the CEF intensities for the \emph{R}SFO compounds, the spin wave excitations of Fe ions in all three \emph{R}SFO compounds were isolated and are compared in Fig. 6(d). The spin wave spectra in all three compounds agree with each other for energies greater than 85 meV. At all other energies, the spin wave spectra of LSFO and PSFO remain similar: there are two energy bands with a gap between 60-80 meV. In NSFO, the low energy spectral weight appears to move to higher energies and fills this portion of the spectrum. The intense peak around 20-30 meV in NSFO is a signal which also exists in both LSFO and PSFO and is likely an artifact due to inaccuracies in subtracting the phonon spectra based upon high angle measurements.

Previous analysis of spin waves in LSFO [6] showed that the spectrum can be modeled adequately for energy transfers above 40 meV where phonon corrections are modest. We base our model on the facts that there are two different kinds of Fe ions, Fe$^{3+}$ and Fe$^{5+}$, and two nearest-neighbor exchange interactions which obey the Goodenough-Kanamori rules for the sign of the exchange; antiferromagnetism between half-filled Fe$^{3+}$-Fe$^{3+}$ pairs and ferromagnetism between half-filled and empty \emph{e}$_g$ orbitals in  Fe$^{3+}$-Fe$^{5+}$ pairs. The \emph{NN} Heisenberg model Hamiltonian is

\begin{equation}
 {\bf{H}} = - J_{AF} \sum_{\langle i, j \rangle} {\bf{S}}_i^{3+} \cdot {\bf{S}}_j^{3+} - J_{F} \sum_{\langle i, j \rangle} {\bf{S}}_i^{3+} \cdot {\bf{S}}_j^{5+} \, ,\label{fiteq}
\end{equation}
where sums are over each pair-type,  ${\bf{S}}_i$ and ${\bf{S}}_j$ represent the spin vector of the \emph{i}th and \emph{j}th iron atom of the type indicated, and sums are over nearest-neighbor pairs with pairwise exchange values (\emph{J}$_{F}$ or \emph{J}$_{AF}$) determined by the charge ordered structure.

Because of the small charge-transfer gap in \emph{R}SFO, significant hybridization exists between Fe and O resulting in some fractions of doped holes residing on the oxygen site. Even considering the presence of doped holes, the exchange between Fe$^{3+}$ and nominal Fe$^{5+}$ ions remains ferromagnetic. In the limit where the holes are on the iron site (corresponding to full Fe$^{5+}$ valence), F exchange occurs between half-filled and empty \emph{e}$_g$ orbitals according to the Goodenough-Kanamori rules. In the limit where a single hole is on the oxygen site, sharing of the spin-polarized oxygen electron also leads to F exchange. For the same reason, the presence of oxygen holes between Fe$^{3+}$ - Fe$^{3+}$ pairs will reduce \emph{J}$_{AF}$ as compared to the parent insulator \emph{R}FO. Furthermore, the spin values of Fe ions are affected by hybridization: if the hole is on the oxygen ion, the Fe oxidation state is lower and the effective spin of the Fe ions would be larger.

\begin{figure*} [tbh]
  \includegraphics[width=0.80\textwidth]{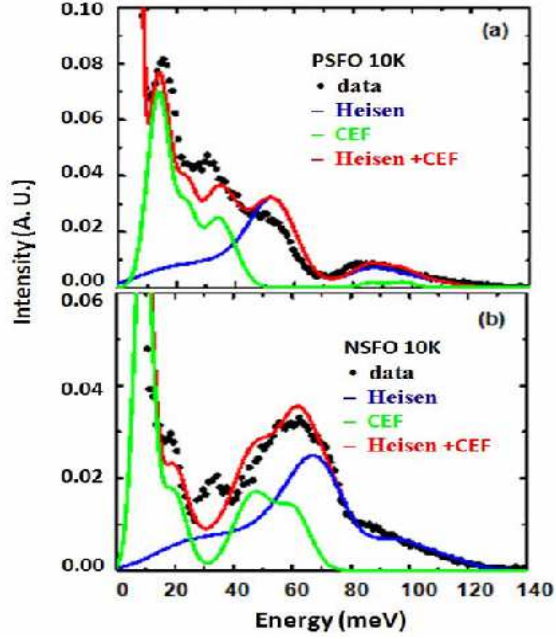}\\
  \caption{(color online) Comparison of \emph{R}SFO (\emph{R} = Pr (a) and Nd (b)) magnetic scattering spectra at \emph{T} = 10 K (dots) and summed for 2$\theta$ 3$^\circ$-30$^\circ$. The Heisenberg model calculation (blue line), the CEF excitations (green line) and the sum of the two calculations (red line) are shown.}
  \label{}
\end{figure*}

Figure 7 shows the fitting results of PSFO and NSFO to the Heisenberg model of Eq. (11). Unlike the simple case of \emph{R}FO, the fitting of \emph{NN} Heisenberg model calculations and CEF excitations in \emph{R}SFO do not show quantitative agreement with the data especially at low energies where phonon corrections are more important. However, as explained in [6] the critical ratio $\mid$\emph{J}$_F$/\emph{J}$_{AF}$$\mid$ can be estimated by the splitting between the upper (> 70 meV) and lower energy (< 60 meV) bands. For LSFO and PSFO, the \emph{NN} Heisenberg model calculations show that these two main spin wave bands originate from F-like/AF-like spin waves that propagate along/between the metal centered domain wall, respectively. A rough estimate of the energy scale for these excitations is,

\begin{equation}
\begin{split}
E_{F}=  3 J_{F} (2S^{3+} + S^{5+}) \, \, \, \, \, \, \, \, \, \, (a), \\
E_{AF}= 3 \mid J_{AF} \mid S^{3+} + 3 J_F S^{5+} \, \, \, \, \, \, (b),
\end{split}\,
\label{fiteq}
\end{equation}
Based on the previous reports on spin magnetic moments for different Fe ions, S$^{3+}$ is very stable and keeps as the value 5/2, while S$^{5+}$ is likely to be larger than the atomic limit of 3/2 due to hybridization and can be adjusted.[22, 23] The final fitting magnetic exchange energies are shown in Table VIII.

\begin{table*} [tph]
\caption{Parameters of the \emph{NN} Heisenberg model for \emph{R}SFO (\emph{R} = La, Pr and Nd) and the size of the charge transfer gap.[5, 24] Parameters of LSFO are from Ref. [6].}
\begin{tabular}{lccc}
\hline
   & $\,  \,  \,  \,  $ La$_{1/3}$Sr$_{2/3}$FeO$_3$ $\,  \,  \,  \,  $  & $\,  \,  \,  \,  $ Pr$_{1/3}$Sr$_{2/3}$FeO$_3$  $\,  \,  \,  \,  $  & $\,  \,  \,  \,  $  Nd$_{1/3}$Sr$_{2/3}$FeO$_3$ $\,  \,  \,  \,  $  \\
\hline
Spin momenta && & \\
$\,  \,  \,  \,  $ \emph{S}$^{3+}$            & 2.5 & 2.5 & 2.5 \\
$\,  \,  \,  \,  $ \emph{S}$^{5+}$             & 2.0 $\pm$ 0.1 & 2.0 $\pm$ 0.1 & 2.0 $\pm$ 0.1 \\
\hline
Magnetic exchange energies && & \\
$\,  \,  \,  \,  $ \emph{J}$_{AF}$ (meV)           & -3.5 $\pm$ 0.2 & -3.5 $\pm$ 0.2 & -5.5 $\pm$ 0.3 \\
$\,  \,  \,  \,  $ \emph{J}$_{F}$ (meV)           & 5.1 $\pm$ 0.1 & 5.1 $\pm$ 0.1 & 5.1 $\pm$ 0.1 \\
$\,  \,  \,  \,  $ ratio ($\mid$\emph{J}$_{F}$/\emph{J}$_{AF}$$\mid$)           & 1.46 $\pm$ 0.05 & 1.46 $\pm$ 0.05 & 0.93 $\pm$ 0.03 \\
\hline
charge transfer gap & & & \\
$\,  \,  \,  \,  $ $\Delta$ (meV)           & 62 & 58 & 85 \\
\hline
\end{tabular}
\label{Table.II.}
\end{table*}

The measured magnetic exchange ratio of \emph{J}$_F$ and $\mid$ \emph{J}$_{AF}$ $\mid$ can be compared to the theoretical predictions of the magnetic exchange mechanism for the CO in \emph{R}SFO (\emph{R} =  La, Pr, and Nd) as shown in Fig. 8. In the magnetic exchange only model proposed by T. Mizokawa, et al.,[7] two charge ordered patterns are considered with alternating charge and spin ordering along either the (111) or (100) directions. Those with the observed pattern of charge ordering along the cubic (111)-direction are stable for large values of the exchange ratio, $\mid$\emph{J}$_F$/\emph{J}$_{AF}$$\mid$ > 1, although the boundary between the two phases at \emph{T} = 10 K depends on the value of \emph{NN} exchange between Fe$^{5+}$ - Fe$^{5+}$ ions (\emph{J}$_{55}$) that is present only in the (hypothetical) (100) charge ordered structure.  According to the Goodenough-Kanamori rules, it is expected that this exchange is weakly AF due to the $\pi$ - bonding of half-filled \emph{t}$_{2g}$ orbitals. In the limit where \emph{J}$_{55}$ $\approx$ 0, the (111) order is stable when $\mid$\emph{J}$_F$ / \emph{J}$_{AF}$$\mid$ > 1/2. A much more conservative estimate of \emph{J}$_{55}$ $\approx$ $\mid$ \emph{J}$_{AF}$ $\mid$ results in the condition  $\mid$\emph{J}$_F$ / \emph{J}$_{AF}$$\mid$ > 1 for the stability of (111) charge order. LSFO and PSFO have exchange ratios that clearly favor the (111) ordering, even in the most conservative estimate for the value of \emph{J}$_{55}$. On the other hand, the exchange ratio for NSFO is slightly less than one. This suggests that the (111)-type charge ordering is less stable in NSFO as compared to LSFO or PSFO. This is consistent with the suppression of \emph{T}$_N$ in the \emph{R}SFO compounds.

\begin{figure}
  \includegraphics[width=0.75\textwidth]{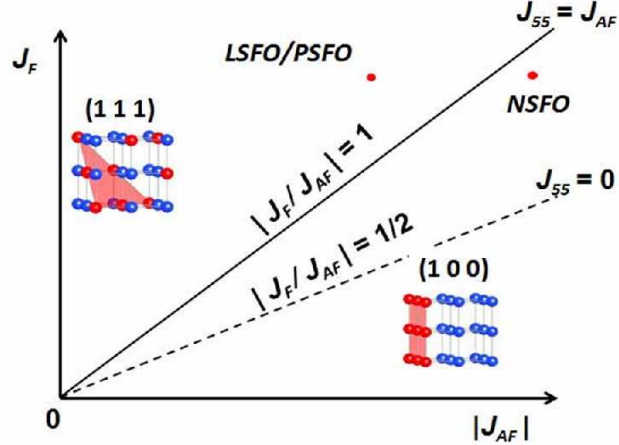}\\
  \caption{(color online) The charged order phase diagram for \emph{R}SFO (\emph{R}=La, Pr, and Nd) as a function of nearest-neighbor magnetic exchange. The (111) and (100) types of charge order are illustrated respectivley in the corresponding portions of the phase diagram.}
  \label{}
\end{figure}

The reduction of the exchange ratio in NSFO largely arises from a $\sim$ 20$\%$ increase of $\mid$\emph{J}$_{AF}$$\mid$ as compared to LSFO and PSFO. The AF exchange in NSFO is comparable to that of the parent \emph{R}FO compounds. Based on Table III and VIII, the increase of $\mid$\emph{J}$_{AF}$$\mid$ in NSFO as compared to LSFO and PSFO could arise from differences in the CT gap and/or the lattice distortion of these compounds. The CT gap of NSFO is larger than those of LSFO or PSFO, hence electrons will be more localized and the magnitude of AF exchange energy of NSFO is expected to increase. In addition, the effect of lattice distortion on the magnetic exchange should also be considered. The effect of the lattice distortions in the \emph{R}FO parent compounds is well understood. The tolerance factor, whose deviation from one indicates the propensity for lattice distortion, decreases from LFO to NFO (see Table II). The larger lattice distortion in NFO results in smaller Fe-O-Fe bond-angles that weakens the AF superexchange. The tolerance factor of the \emph{R}SFO compounds is closer to 1 than the parent \emph{R}FO compound and varies only weakly throughout the \emph{R}SFO series. Therefore, the large change in \emph{J}$_{AF}$ is unlikely to arise from an average change of the doping induced structural distortions.

\section*{V) Conclusion}

Using inelastic neutron scattering, we determined that the similar spin wave spectra of LSFO and PSFO consist of two energy bands separated by a large energy gap, while the two bands merge into one in NSFO. The full magnetic bandwidth is determined mainly by the ferromagnetic exchange energy, \emph{J}$_{F}$, between Fe$^{3+}$ and Fe$^{5+}$ ions, and is found to be similar for the different \emph{R}SFO compounds. The AF exchange energies between Fe$^{3+}$ ions, $\mid$\emph{J}$_{AF}$$\mid$, which controls the splitting of the upper and lower magnetic bands are more sensitive to \emph{R} substitution. We determine \emph{J}$_{F}$ and $\mid$\emph{J}$_{AF}$$\mid$ by comparison to a Heisenberg model. The ratio of these exchanges is an indicator of the role that magnetism plays in the formation of the charge ordered state. While LSFO and PSFO are in the regime where magnetic exchange can stabilize the charge ordered state, the case for NSFO is not as clear. The much lower exchange ratio in NSFO may come from the increase of the charge transfer gap that is caused by the smaller Nd ion.

This work was supported by the Division of Materials Sciences and Engineering, Office of Basic Energy Sciences, U.S. Department of Energy. Ames Laboratory is operated for the U.S. Department of Energy by Iowa State University under Contract No. DE-AC02-07CH11358. The work at the Spallation Neutron Source, Oak Ridge National Laboratory (ORNL), was sponsored by the Scientific User Facilities Division, Office of Basic Energy Sciences, U.S. Department of Energy (U.S. DOE). ORNL is operated by UT-Battelle, LLC for the U.S. DOE under Contract No. DE-AC05-00OR22725. The work has benefited from the use of the Los Alamos Neutron Science Center at Los Alamos National Laboratory. LANSCE is funded by the U.S. Department of Energy under Contract No. W-7405-ENG-36.

JM, JQY, and RJM thank S. Chang, R. W. McCallum, D. C. Johnston and P. C. Canfield for the help in sample synthesis and characterization.

\bibliographystyle{apsrev}

\end{document}